\documentstyle[twocolumn,epsfig,aps]{revtex}
\draft

\begin{document} 
\twocolumn[\hsize\textwidth\columnwidth\hsize\csname
@twocolumnfalse\endcsname \title{Conserved Ordering Dynamics of Heisenberg Spins
with Torque} 
\author{Jayajit Das$^{1,2}$\cite{JAY} and Madan Rao$^2$\cite{MAD}}

\address{$^1$Institute of Mathematical Sciences, Taramani, Chennai
600113,
India\\
$^2$Raman Research Institute, C.V. Raman Avenue,
Sadashivanagar, Bangalore 560080,
India}

\date{\today}

\maketitle

\begin{abstract}

We show that a torque induced by the local molecular field drives the
zero-temperature ordering dynamics of a conserved Heisenberg magnet to a
new fixed point, characterised by exponents $z=2$ and $\lambda \approx
5.15$. Numerical solutions of the Langevin equation indicate that theories
using a Gaussian closure are {\it inconsistent} even when the torque is
absent.  The torque is relevant even for quenches to $T_c$, with exponents
$z=4-\varepsilon/2$ and $\lambda = d$ (where $\varepsilon = 6-d$).  
Indeed $\lambda$ is always equal to $d$ for quenches to $T_c$
whenever the order parameter is conserved.

\end{abstract}

 ] \vskip1.0in

When a magnet is quenched from its disordered high temperature phase to
its ordered configuration at low temperatures, the slow annealing of
``defects'' separating competing domains, makes the dynamics very
slow. The system organizes itself into a self similar
spatial distribution of domains characterised by a single diverging length
scale which typically grows algebraically in time $L(t) \sim t^{1/z}$.
This spatial distribution of domains is reflected in the scaling behaviour
of the equal-time correlation function $C(r,t) \sim f(r/L(t))$.  The
autocorrelation function, $C(0, 0\,; 0, t) \sim L(t)^{-\lambda}$ is a measure
of the memory of the initial configurations. The exponents $z$ and
$\lambda$ and the scaling function $f(x)$ characterise the dynamical
universality classes at the zero temperature fixed point (ZFP)\cite{BRAY}.

There has been a trend in recent years to compare the theories of phase
ordering dynamics with numerical simulations of Langevin equations. 
Comparison with experimental systems, such as magnets, binary fluids or
binary alloys have to take into account the various `real' features that
might be relevant to its late time dynamics. For instance, theories of
binary fluids have to include effects of hydrodynamics, while those of
binary alloys have to incorporate elastic and hydrodynamic effects. In the
same vein, any comparison with the dynamics in real magnets has to include
the effects of the torque induced by the local molecular field. 

In this article, we study the conserved phase ordering dynamics of a
Heisenberg magnet in three dimensions in the presence of a torque (the
corresponding nonconserved dynamics has been studied in Ref.\ \cite{DAS}). 
For quenches to $T=0$, our Langevin simulation conclusively shows that the
torque drives the dynamics to a new ZFP characterised by $z=2$ and
$\lambda \approx 5.15$. This is confirmed by simple scaling arguments,
which in addition gives the crossover time and the crossover exponent. 
Further we demonstrate that the approximate theories of conserved dynamics
for vector order parameters based on the Gaussian closure
scheme\cite{MAZENKO}, are internally inconsistent even when the torque is
absent, contrary to what has been assumed in the
literature\cite{BRAY}.
 
We investigate the dynamics following a quench to the critical point $T_c$
and show that the torque is relevant at the Wilson-Fisher fixed point.
Using a diagrammatic perturbation theory, we show that $z=4-\varepsilon/2$
at this new fixed point (where $\varepsilon = 6-d$). We show to all orders
in a perturbative expansion, that the autocorrelation exponent $\lambda =
d$.  This last result is true whenever the order parameter is conserved
and can be understood from very general arguments\cite{SM}. 

The Heisenberg model in 3-dim is described by a vector order parameter
$\vec\phi({\bf r},t)$ and a free-energy functional,

\begin{equation}
F\left[\vec \phi\right] = \int d\,{\bf r} \left[ \frac{1}{2}\left(\nabla
\vec\phi\right)^2 + \frac{r_0}{2} \vec\phi\cdot\vec\phi + \frac{u}{4}
\left(\vec\phi\cdot\vec\phi\right)^2\right]\,\,. 
\label{eq:free}
\end{equation}

This leads to the following dynamical equation (in dimensionless
quantities), describing zero temperature quenches of the conserved
Heisenberg model \cite{DAS},

\begin{equation}
\frac{\partial \vec{\phi}}{\partial t} = \nabla^2\left(\nabla^2 \vec{\phi} \, +
\, \vec{\phi} - \, (\vec{\phi} \cdot \vec{\phi})\, \vec{\phi}\right) \, + \,g
\, \left(\vec{\phi} \times \nabla^2\vec{\phi}\right) \, . 
\label{eq:dyeq}
\end{equation} 

The first term on the right side is the dissipative force, while the
second term is the torque generated by the local molecular field.  The
dimensionless parameter $g \sim {\Omega}/{\Gamma} $ is the ratio of
the precession frequency to the relaxation rate and is in the range
$10^{-3} - 10$ for real magnets. 

We discretize Eq.\ (\ref{eq:dyeq}) on a simple cubic lattice (with size
$N$ ranging from $50^3$ to $60^3$) adopting an Euler scheme for the
derivatives.  We include the next-nearest neighbour contributions in
calculating the second and higher order spatial derivatives which reduces
lattice anisotropy and gives better numerical stability \cite{SR}. The
space and time intervals have been chosen to be $\triangle x = 2.5$ and
$\triangle t = 0.2$ which lead to stable results for the resulting coupled
map. We have checked that slight variations of $\triangle x$ and
$\triangle t$ do not change the results.  Throughout our simulation we
have used periodic boundary conditions.  We prepare the system initially
in the paramagnetic phase, where $\vec \phi$ is distributed uniformly with
zero mean (angular brackets denote an average over this distribution and
over space). In our simulations we find that averaging over $5$ initial
uncorrelated configurations gives us clean results (the statistical
errors in $C(r,t)$ are at most $3\%$). 

We compute the equal-time correlator $C(r, t) \equiv \langle {\vec
\phi}(r, t) \cdot {\vec \phi}(0,t) \rangle$, the autocorrelation function,
$C(0, t_{1}=0, t_{2}) \equiv \langle \,{\vec \phi}(0, 0) \cdot {\vec
\phi}(0, t_2) \rangle$ and the energy density.  Figure\,$1$ is a scaling
plot of $C(r,t)$ versus $r/L(t)$ for various values of the parameter $g$,
where $L(t)$ is extracted from the first zero of $C(r,t)$. Note that the
scaling function for $g=0$ is very different from those for $g>0$\,;  
further the $g>0$ scaling functions do not seem to depend on the value of
$g$. This suggests that the torque term drives the dynamics to a new ZFP.  
This is also revealed in the values of the dynamical exponent $z$. In
Fig.\ 2, a plot of $L(t)$ versus $t$ gives the expected value of $z=4$
when $g=0$. For $g>0$, we see a distinct crossover from $z=4$ when
$t<t_c(g)$ to $z=2$ when $t>t_c(g)$. The crossover time $t_c(g)$ decreases
with increasing $g$.  The same $z$ exponent and crossover are obtained
from the scaling behaviour of the energy density, $\varepsilon =
\frac{1}{V}\, \int d {\bf r} \,\langle \,(\,\nabla \vec \phi({\bf r},
t)\,)^2 \,\rangle \sim L(t)^{-2}$.

\begin{figure}
\centerline{\epsfig{figure=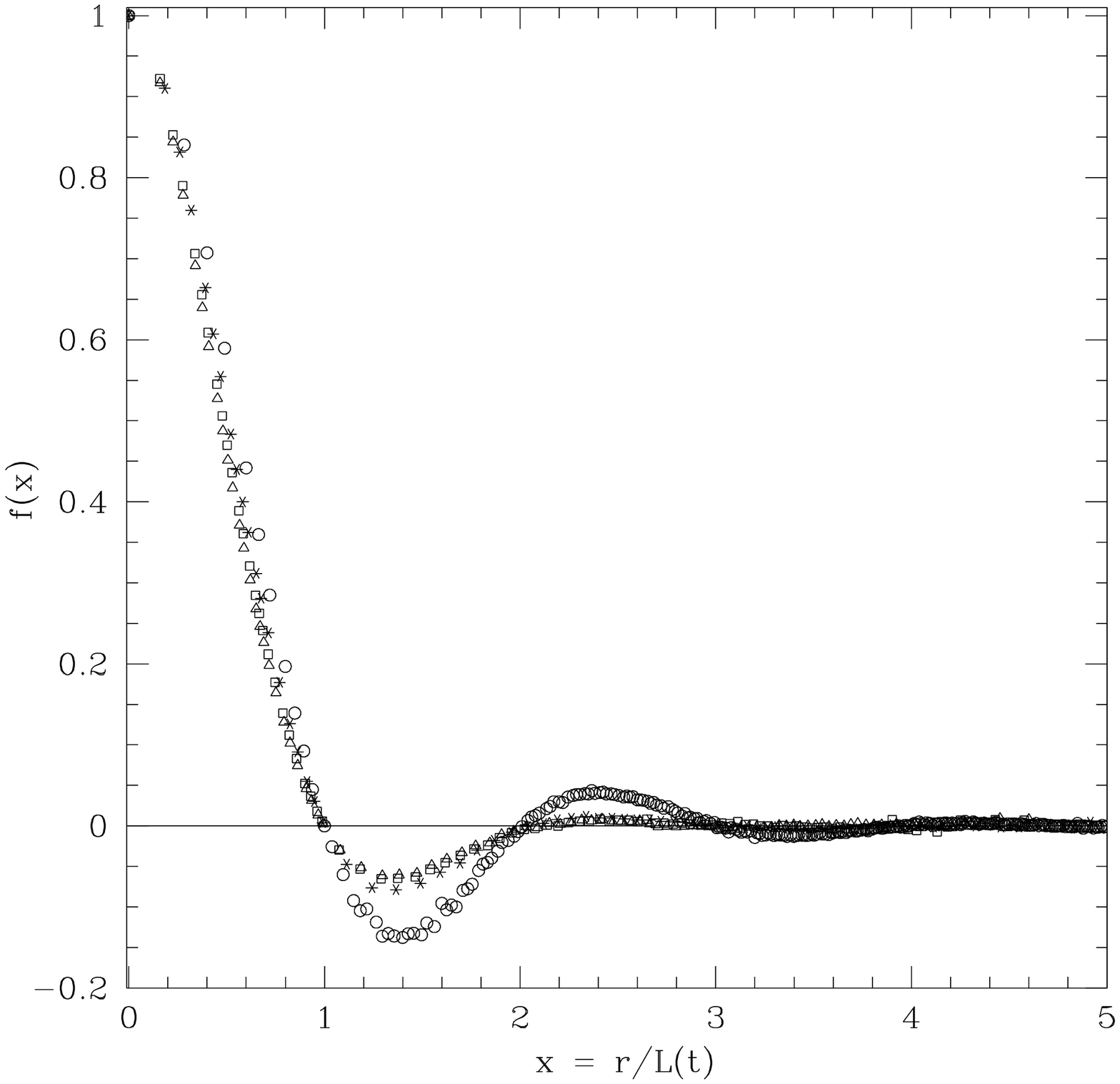,width=7.0cm,height=7.0cm}}
\end{figure} 
FIG. 1. Scaling plot of the equal time correlation function for $N=50^3$. 
Observe the crossover in the scaling function $f(x)$ as $g$
changes from $g=0(\circ)$ to nonzero values,
$g=0.1(\ast),\,g=0.3(\triangle)\,$ and $g=0.5(\Box)$. \\

The $\lambda$ exponent extracted from the autocorrelation function
$C(0,0,t)$ also suggests a crossover to a new fixed point. The numerical
determination of $\lambda$ is known to be subject to large
errors\cite{DAS}, and so we have to go to very late times and hence
large system sizes.  For the $60^3$ lattice we take data when $900 \leq t
\leq 9000$ and average over $7$ initial configurations. This time range is
well within the scaling regime for $C(r,t)$ when $g =
0,\,0.1,\,0.2,\,0.3$. We compare the data for each $g$ to
$A(t+t_0)^{-\lambda/z}$ with $A$, $t_0$ and $\lambda$ as fit parameters.
The data show a crossover to a new ZFP beyond a time $t_c(g)$. 
We quote the values obtained for $t > t_c(g)$ --- $\lambda(g=0) \approx 2.2$, 
$\lambda(g=0.2) \approx 5.10$,
$ \lambda(g=0.3) \approx 5.17$. 

We provide a simple scaling argument to understand some of the results
quoted above. On restoring appropriate dimensions, the dynamical equation
Eq.(\ref{eq:dyeq}) can be rewritten as a continuity equation,
$\partial{\vec\phi({\bf r},t)}/{\partial t } = - {\bf \nabla }\cdot{\vec j
}$ where the ``spin current''

\begin{equation}
 \vec{j}_{\alpha} = -\Gamma\left({\bf \nabla }\frac{\delta F[\vec \phi]}{\delta
\phi_{\alpha}} +
\frac{\Omega}{\Gamma}\epsilon_{\alpha\beta\gamma}\phi_{\beta}\nabla\phi_{
\gamma}\right) \, .
\label{eq:current} 
\end{equation} 

Using a dimensional analysis where we replace $j_{\alpha}$ by the
`velocity' $dL/dt$, we find

\begin{equation} 
\frac{dL}{dt} = \Gamma\frac{\sigma}{L^3} +
\Omega\frac{\sigma M_{0}}{L}\, ,
\label{eq:dimension} 
\end{equation}
where $M_0$, $\sigma$ and $\Gamma^{-1}$ are the equilibrium
magnetisation, surface tension and spin mobility respectively.
Beyond a crossover time given by $t_c(g) \sim (\Gamma/M_{0} \Omega)^2 \sim
1/g^2$, simple dimension counting shows that the dynamics crosses over
from $z = 4$ to $z = 2$ in conformity with our numerical simulation.

\begin{figure}
\centerline{\epsfig{figure=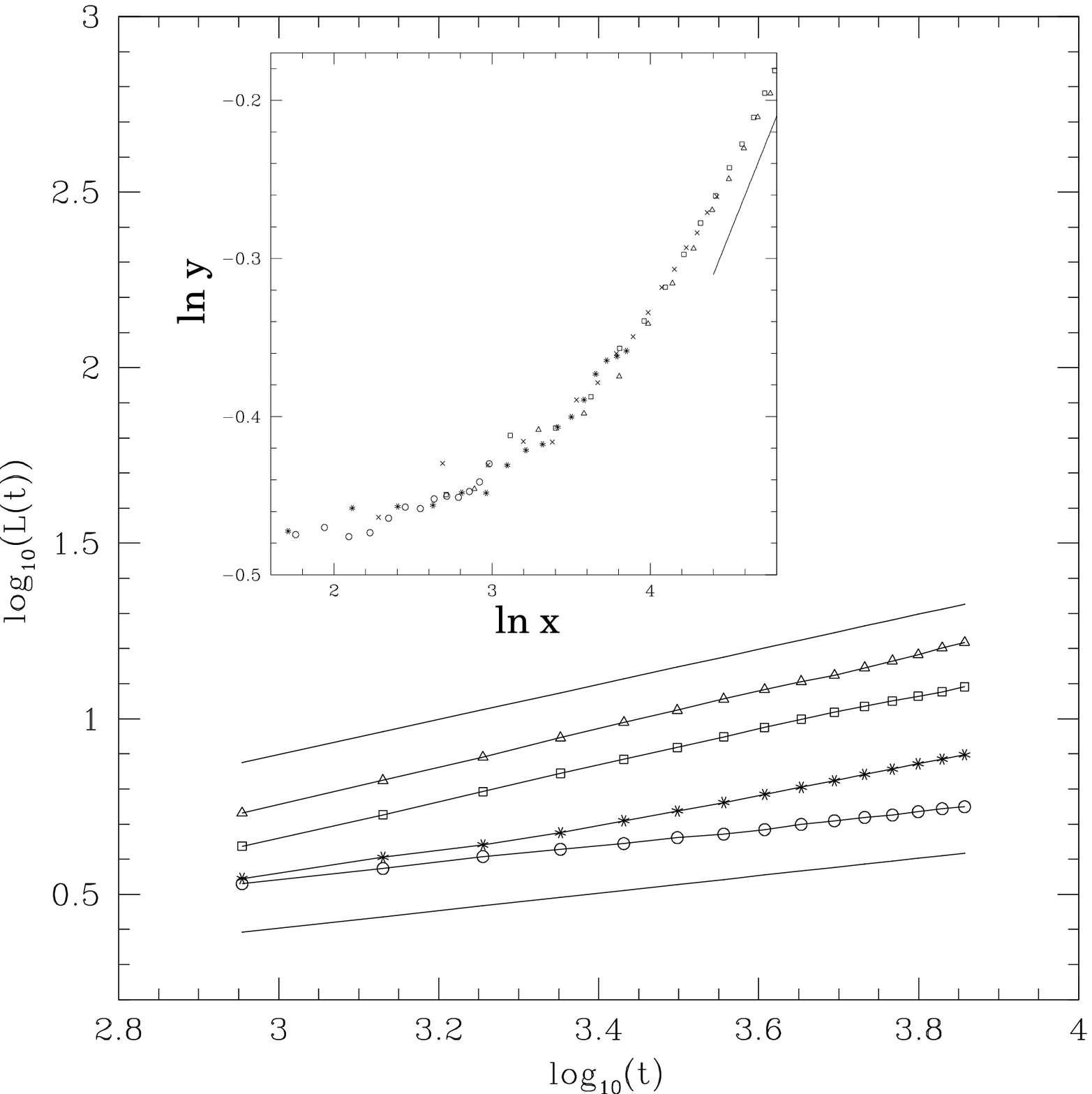,width=7.0cm,height=7.0cm}}
\end{figure} 
FIG. 2. log-log plot of $L(t)$ versus $t$. 
At $g=0\,(\circ)$, $z=4$ (line of slope $0.25$ 
drawn 
for comparison). At other values of $g\, (g=0.1(\ast),\,0.3(\Box),\,
0.5(\triangle))$, $z$ crosses over from 
$4$ to $2$ (line of slope $0.5$
drawn for comparison). Inset shows a scaling plot
of $y=L(t,g)/t^{1/4}$ versus $x = tg^{1.7}$ (see text), for
$g=0.03(\circ),\,0.05(\ast),\,0.07(\times),\,0.09(\Box),\,0.10(\triangle)$.
The scaling function asymptotes 
to a line of slope $0.25$ (solid line) as $x\rightarrow \infty$.
\\

The
$g$ dependence of the crossover time $t_c$ is also borne out by our
numerics (Fig.\ 2(inset)) as we demonstrate below.
Equation(\ref{eq:dimension}) suggests the scaling ansatz $L(t,g) =
t^{\alpha} s(tg^{\phi})$, valid for all $g$. This scaling form is governed
by the $g=0$ ZFP, and so the scaling function $s(x)$ should asymptote to a
$g$-independent constant $s_0$ as $x \to 0$. This implies that $\alpha =
1/4$. On the other hand, as $x\to \infty$, $s(x) \sim x^{1/4}$. If the
above proposal is true, then the data for $L(t,g)/t^{1/4}$ versus
$tg^{\phi}$ should collapse onto a scaling curve for an appropriate value
of $\phi$. We find that $\phi \approx 1.7$. Thus the crossover time,
obtained when $tg^{\phi}=1$, goes as $t_c \sim g^{-1.7}$, close to our
earlier estimate. Note that our numerical estimate of the crossover
exponent can be improved by introducing finite-time shift factors.

Can the above results be understood within the class of approximate
theories based on the gaussian closure approximation\cite{MAZENKO,BRAY}?
The gaussian closure method consists of trading the order parameter ${\vec
\phi} ({\bf r},t)$ which is singular at defect sites, for an everywhere
smooth field ${\vec m}({\bf r},t)$, defined by a nonlinear transformation,
$ {\vec \phi}({\bf r}, t) = {\vec \sigma}\left({\vec m}({\bf r}
,t)\right)\, $. Correlation functions are calculated making the single
assumption that each component of ${\vec m}({\bf r},t)$ is an independent
gaussian field with zero mean at all times. The equal time correlation
function takes the form \cite{MAZENKO}
\begin{equation}
C(r,t)=\frac{3\gamma}{2\pi}\left[B\left(2,\frac{1}{2}\right)\right]
^{2} 
{}_2F_{1}\left(\frac{1}{2},\frac{1}{2},\frac{5}{2};\gamma^2\right) 
\label{eq:hyper}
\end{equation}
where $B(x,y)$ and ${}_2F_{1}(a,b,c;z)$ are the Beta and hypergeometric
functions respectively and $\gamma(r,t)=\langle{\vec m}({\bf r}+{\bf
x},t)\cdot{\vec m}({\bf x},t)\rangle/\langle m({\bf x},t)^2\rangle $. This
scheme which works remarkably well for nonconserved systems \cite{DAS}
fails to give consistent results for our conserved dynamics, as we show, 
using a criterion developed by Yeung et. al. \cite{CHUCK}
for a conserved scalar order parameter.

\begin{figure}
\centerline{\epsfig{figure=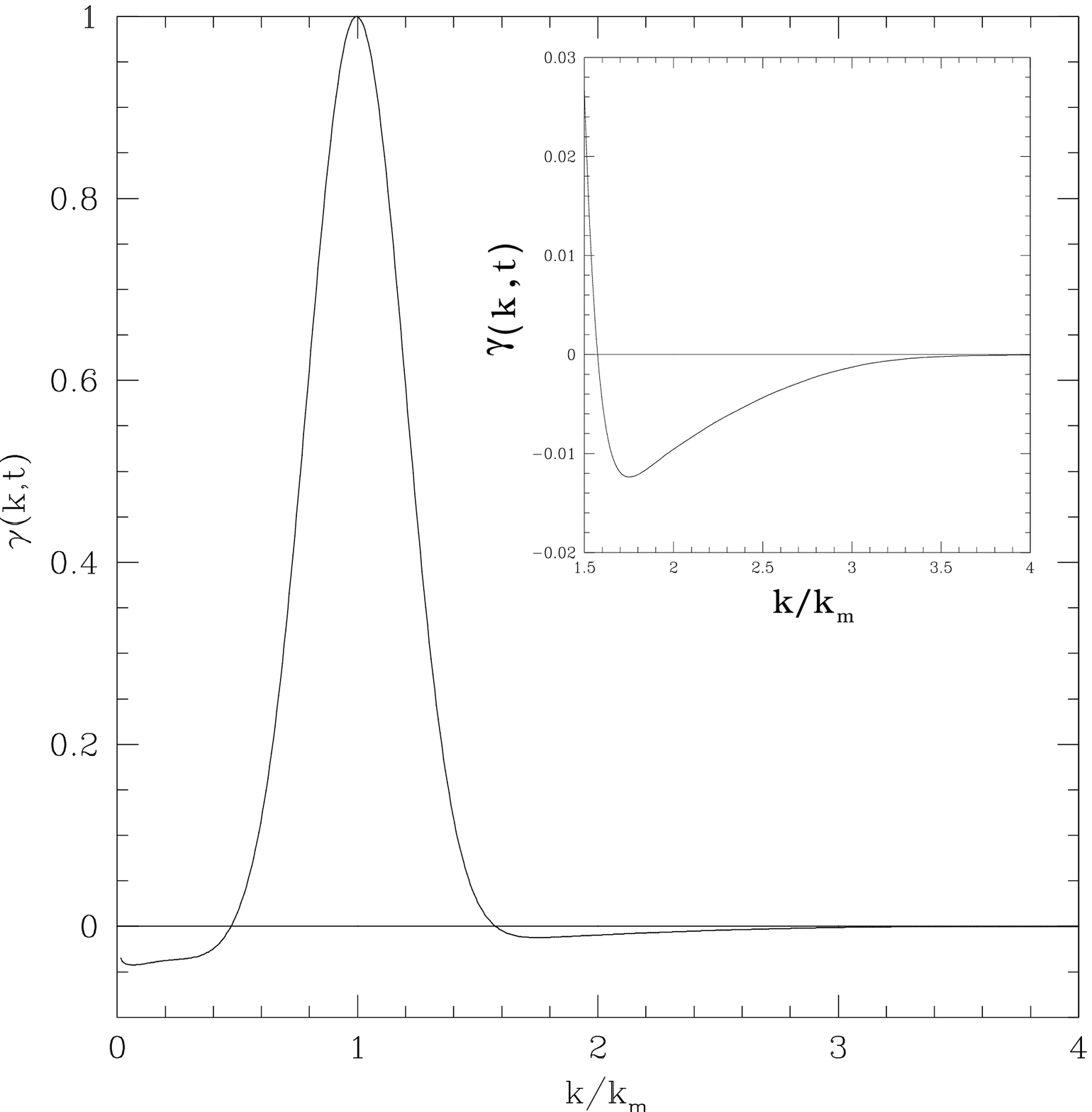,width=7.0cm,height=7.0cm}}
\end{figure} 

FIG. 3. The spectral density $\gamma(k,t)$ at $t=3600$
becomes negative for $0\leq k/k_{m}<0.5$ and for $1.5<k/k_{m}<3.0$
(inset).

\vskip.3cm

We numerically evaluate the spectral density, the fourier transform of
$\gamma(r,t)$. This is prone to numerical errors because of statistical
errors in our computed $C(r,t)$. For instance, a numerical integration of
$\int d{\bf r} C(r,t)$ provides a nonzero value whereas it should be
identically zero because of the conservation law. This is reflected in
large errors in $\gamma({\bf k},t)$ at small ${\bf k}$.
We therefore adopt the following procedure.  We fit a function
$C_{f}(x)$\cite{FIT} to the equal time correlation function $C(r,t)$ and
use this to extract $\gamma(k,t)$ from the Eq(\ref{eq:hyper}). We observe
(Fig.\,3) that the spectral density, which should be a strictly positive
function of its arguments, becomes negative for $k/k_{m} < 0.5 $
($\gamma(k,t)$ is peaked at $k_{m}$)  and in the range $1.5<k/k_{m}<3.0$ !
This demonstration highlights the intrinsic inconsistency of the gaussian
approach for conserved vector order parameters.

We investigate the dynamics of the order parameter quenched to the
critical point $T_c$. At the Wilson-Fisher fixed point,
$u^{*}=(8/11)\pi^2\epsilon$  ($\epsilon = 4 - d$),
the scaling dimension of $g$ is 
$d/2 + 1 - z + \eta/2$, where $z=4-\eta$ and $\eta = (5/242) \epsilon^2$
implying that the torque $g$ is relevant when $d<6$.

We calculate the dynamical exponents $z$ and $\lambda$ at this new
fixed point. This is done using a diagrammatic perturbation theory
within the Martin-Siggia-Rose (MSR) formalism \cite{JANS}. For our problem,
the MSR generating functional is,
${\cal Z}[\vec{h},\vec{\tilde{h}}] = \int{\cal
D}(\vec{\tilde{\phi}}){\cal
D}(\vec{\phi})\exp\,(-J[\vec{\phi},\vec{\tilde{\phi}}]-H_{0}[\vec{\phi}_{0}]
+\int_{0}^{\infty} dt \int d{\bf k}
(\vec{\tilde{{h}}}_{\bf k}\cdot\vec{\tilde{\phi}}_{-\bf k}+\vec{h}_{\bf
k}\cdot\vec{\phi}_{-\bf k}) $
where the MSR action is 

\begin{eqnarray}
\lefteqn{J[\vec{\phi},\vec{\tilde{\phi}}] = \int_{0}^{\infty}dt\int d{\bf k}
\bigg\{\vec{\tilde{\phi}}_{\bf
k}\cdot\bigg[\partial_{t}\vec{\phi}_{\bf k}+\Gamma k^2\frac{\delta
F[\vec{\phi}]}{\delta \vec{\phi}_{-\bf k}}
} \nonumber \\
                                      &  & +\int d{\bf
k}_1\bigg(\frac{g\Gamma}{2}(k_{1}^{2}-({\bf k}-{\bf k}_{1})^2)
\vec{\phi}_{{\bf k}_1}\times\vec{\phi}_{{\bf k}-{\bf k}_1}\bigg)
\bigg] \nonumber \\ 
                                     &  &
 -\Gamma k^2\vec{\tilde{\phi}}_{\bf  k}\cdot\vec{\tilde{\phi}}_{-\bf
k}\bigg\} \,\, ,
\end{eqnarray}

and $H_0$ denotes initial distribution (gaussian with width
$\tau^{-1}_{0}$ and spatially
uncorrelated) of the order parameter
$H_{0} = \int d{\bf k}\frac{\tau_{0}}{2}(\vec{\phi}_{\bf
k}(0)\cdot\vec{\phi}_{-\bf k}(0))$ \cite{JANS}. 

Power counting reveals the presence of
two different upper critical dimensions coming from the 
quatric term ($d_c^{u}=4$) and the cubic torque term ($d_c^{g}=6$) in 
the action $J$. This implies we have to
evaluate the fixed points and exponents in a double power series expansion
in $\epsilon=4-d$ and $\varepsilon=6-d$ \cite{HAL}. 

The unperturbed correlation $C^{0}_{\bf k}(t_1,t_2)$ and response
$G^{0}_{\bf k}(t_1,t_2)$ functions, and the bare $u$ and $g$ vertices are
shown in Fig.\ 4.  Again power counting shows that at $d=3$, our
perturbation expansion does not generate additional terms other than those
already contained in $J$, i.e. the theory is renormalizable. However the
perturbation theory gives rise to ultraviolet divergences which can be
removed by adding counter-terms to the action.

To remove these divergences,
we introduce renormalization factors
(superscripts $R$ and $B$ denote renormalized 
and bare quantities respectively),
$\vec{\tilde\phi}^{R}_{\bf k}(0)= (\tilde ZZ_0)^{-1/2}
\vec{\tilde\phi}^{B}_{\bf k}(0)$, $ \vec\phi^{R}_{\bf
k}(t) = Z^{-1/2} \vec\phi^{B}_{\bf k}(t)$,
$\vec{\tilde \phi}^{R}_{\bf k}(t)  =   {\tilde Z}^{-1/2}
\vec{\tilde \phi}^{B}_{\bf k}(t) $,
$u^{R} =  Z^{-1}_{u}u^{B}$,
$g^{R} = Z^{-1}_{g}g^{B}$,
$\Gamma^{R} = Z^{-1}_{\Gamma}\Gamma^{B}$ and 
$\tau^{R}_{0} = Z^{-1}_{\tau_0} \tau^{B}_{0} $.

Since the dynamics obeys detailed balance, the
renormalization factors $Z$ and $Z_{u}$ are the same as in
statics. Further the conservation of the order parameter forces $Z\tilde Z=1$ to all
orders.

The new fixed point is given by the zeroes of the $\beta$ functions of the
theory. The $\beta$ functions, calculated from the $Z$ factors, 
get contributions from all diagrams
containing the primitively divergent diagrams $\Gamma^{(2)}_{\phi 
\tilde\phi}$, $\Gamma^{(3)}_{\phi \phi \tilde\phi}$ and 
$\Gamma^{(4)}_{\phi \phi \phi \tilde\phi}$ (Fig. 4).

The new fixed point, to one loop, is given by
$g^{*}=\pm \sqrt{192\pi^3\varepsilon} + {\cal O}(\varepsilon^{3/2})$, 
$u^{*}=(8/11)\pi^2\epsilon + {\cal O}(\epsilon^2)$ (note
$u^{*}$ does not change from its WF value to all loops) and the dynamical 
exponent $z=4-\varepsilon/2+{\cal O}(\epsilon^2)$ \cite{HAL}. 

\begin{figure}
\centerline{\epsfig{figure=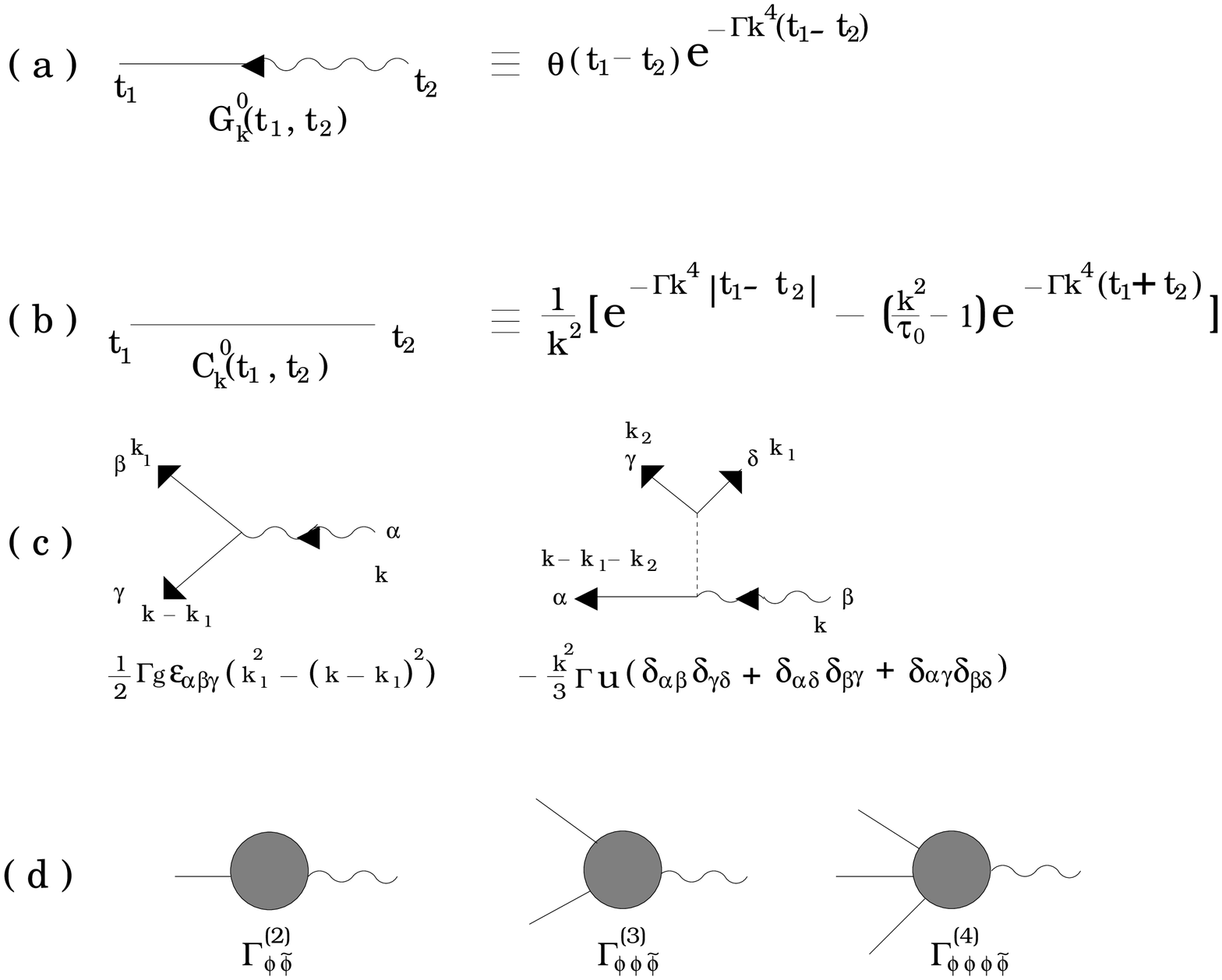,width=7.0cm,height=7.0cm}}
\end{figure} 

FIG.4. Unperturbed (a) response function $G^{0}_{\bf k}$, (b)
correlation function $C^{0}_{\bf k}$, and the (c) two bare vertices $u$ and $g$.
Wavy and straight lines represent the $\vec{\tilde{\phi}}_{\bf
k}(t)$ and $\vec{\phi}_{\bf k}(t)$ fields respectively. (d) Primitively
divergent diagrams $\Gamma^{(2)}_{\phi 
\tilde\phi}$, $\Gamma^{(3)}_{\phi \phi \tilde\phi}$ and 
$\Gamma^{(4)}_{\phi \phi \phi \tilde\phi}$.

\vskip.3cm

The $\lambda$ exponent can be computed from the response 
function $G_{\bf k}(t,0) \equiv \langle {\vec{\tilde
\phi}}_{\bf k}(0) \cdot {\vec \phi}_{-\bf k}(t) \rangle$ since this is 
equal to the autocorrelation function
${\tau_0}^{-1} \langle {\vec \phi}_{\bf k}(t) \cdot {\vec \phi}_{-\bf
k}(0) \rangle$, as can be seen from the first term in $J$ on integrating
by parts. The response function gets renormalized by

\begin{equation}
G^{R}_{k}(t,0)=Z_0^{-1/2} G^{B}_{k}(t,0)\,.
\end{equation}

The divergent contributions to $G_{B}$ could come from two sources. Each
term in the double perturbation series could contain the 
primitively divergent subdiagrams
$\Gamma^{(2)}$, $\Gamma^{(3)}$ or $\Gamma^{(4)}$, which we replace by 
their renormalized counterparts.
The other divergent contribution could arise
from the primitive divergences of the 1-particle reducible vertex function
$\Gamma^{(2)}({\bf k}, t, 0)$, defined by $G_{\bf k}(t,0) \equiv \int G_{\bf
k}(t-t')\,\Gamma^{(2)}({\bf k}, t', 0)\, d\,t'$. The superficial 
divergence of the diagrams contributing to $G_{\bf k}(t,0)$ is
$D=V_u(d-4)+\frac{V_g}{2}(d-6)-2 $ (where $V_u\,(V_g)$ is the number of $u$
($g$) vertices respectively). This is always negative for $d \leq 6$. This 
implies 
that $G^{B}_{\bf k}(t,0)$ does not get renormalized and $Z_0=1$.
Consequently
$\lambda$ stays at its mean-field value of $d$ for
this conserved Heisenberg dynamics both with and without the torque 
\cite{NOTE}. 
 
We have shown that the inclusion of a torque to the ordering dynamics of a
conserved Heisenberg magnet, is relevant both for quenches to $T=0$ and
$T=T_c$. The new zero temperature fixed point is characterised by
exponents $z=2$ and $\lambda \approx 5.15$. We have provided scaling
arguments to understand these exponents and the crossover. We have shown
that the class of approximate theories based on the Gaussian closure
scheme which had been constructed to understand this zero temperature
conserved dynamics, are inconsistent even when the torque is absent.  
On the other hand, the new critical fixed point is characterised by
exponents $z=4-\varepsilon/2$ and $\lambda = d$ (where $\varepsilon =
6-d$).  Indeed $\lambda$ is always equal to $d$ for quenches to the
critical point whenever the order parameter is conserved.

We thank Gautam Menon for a critical reading of the manuscript and
Deepak Dhar for an interesting discussion.

\end{document}